\documentclass[prd, preprint, longbibliography, 11pt]{revtex4-1}

\usepackage{amsmath}
\usepackage{amssymb}
\usepackage{setspace}
\usepackage{graphicx}
\usepackage{natbib}
\usepackage{float}
\usepackage{tensor}
\usepackage[utf8]{inputenc}
\usepackage{amsfonts}
\usepackage{braket}
\usepackage{esint}
\usepackage{breqn}
\usepackage{IEEEtrantools}

\begin{document}
 
 %

\begin{center}
{ \large \bf Nature does not play Dice at the Planck scale}

\smallskip

\vskip 0.1 in

{\large{\bf Tejinder P.  Singh }}

\smallskip

{\it Tata Institute of Fundamental Research,}
{\it Homi Bhabha Road, Mumbai 400005, India}\\
\smallskip
 {\tt tpsingh@tifr.res.in}

\end{center}
\vskip 1 in

\centerline{\bf ABSTRACT}
\smallskip

\noindent  We start from classical general relativity coupled to matter fields. Each configuration variable and its conjugate momentum, as also space-time points, are raised to the status of matrices [equivalently operators]. These matrices obey a deterministic Lagrangian dynamics at the Planck scale. By coarse-graining this matrix dynamics over time intervals much larger than Planck time, one derives quantum theory as a low energy emergent approximation. If a sufficiently large number of degrees of freedom get entangled, spontaneous localisation takes place, leading to the emergence of classical space-time geometry and a classical universe. In our theory, dark energy is shown to be a large-scale quantum gravitational phenomenon. Quantum indeterminism is not fundamental, but results from our not probing physics at the Planck scale. 

\vskip 1 in

\centerline{April 1, 2020}

\bigskip

\centerline{Essay written for the Gravity Research Foundation 2020 Awards for Essays on Gravitation}

\newpage

\noindent Is quantum indeterminism a fundamental property of nature, or is it a consequence of coarse-graining an underlying deterministic theory at the Planck scale? We show the latter to be the case, just as the apparent random motion of a grain of pollen in a glass of water is a consequence of coarse-graining the deterministic motion of water molecules at the microscopic scale. As a by-product of our theory, we show that the currently observed dark energy in the universe is an infra-red quantum gravitational phenomenon.

We begin from a classical theory which has matter fields, Yang-Mills fields, and gravity, living on a Riemannian space-time. In principle we know how to write down the Lagrangian density and action for such a theory. Classical equations of motion follow from extremising this action, which gives the Lagrange equations. A quantum field theory can be constructed from here by imposing the canonical quantum commutation relations, and employing the Heisenberg equations of motion for the quantum operator degrees of freedom. We suggest that this is not the correct theory of nature at the Planck scale. Because this theory cannot dynamically explain the absence of superpositions of space-time geometries in the classical limit. Nor can it dynamically explain why macroscopic position superpositions are absent in the world of classical objects. 

To address these shortcomings of quantum field theory, we propose to start from classical field theory as follows, and construct a deterministic dynamics at the Planck scale. From this deterministic theory at the Planck scale, an emergent low-energy quantum field theory, along with quantum indeterminism, will be derived by coarse-graining over length scales much larger than Planck scale. The degrees of freedom at the Planck scale are averaged over, and play the same role as the molecular motion of water molecules, when we try to understand the motion of a pollen grain. Between every two kicks, the motion is deterministic, whereas from one kick to the next it is random. This is analogous to unitary Schrodinger evolution, followed by non-unitary wave-vector reduction.

Every classical c-number degree of freedom, and its corresponding canonical momentum, is to be replaced by an operator, equivalently a matrix. This is just as is done during quantisation, except that now we will not impose quantum commutation relations. All commutation relations will be allowed to be arbitrary, time-dependent functions, with their time-dependence now determined by the dynamics itself. The Lagrangian of the theory will be constructed by taking the matrix trace of the operator polynomial that results from replacing the configuration variables by the corresponding matrices. Thus for example the harmonic oscillator Lagrangian and action will map to a matrix dynamics as follows:
\begin{equation}
S = \int dt \; [\dot{q}^2 - q^2] \quad \longrightarrow \quad S = \int dt \; Tr [\dot{\bf q}^2 - {\bf q}^2] 
\end{equation}
giving the operator equation of motion $\ddot{\bf q}=-{\bf q}$ after using the `trace derivative' to obtain the new Lagrange equations of motion. This is the essence of Adler's theory of trace dynamics \cite{Adler:04}, which we employ, along with Connes' non-commutative geometry programme \cite{Connes2000}, to construct the deterministic matrix dynamics on the Planck scale
\cite{Maithresh2019, maithresh2019b}.

Given a Riemannian space-time, the Dirac operator $D_B \equiv i\gamma^{\mu}\partial_{\mu}$ on this space-time plays a central role in our theory, because of a theorem in geometry. The heat-kernel expansion of $Tr [L_P^2 D_B^2]$, when truncated at the second order in an expansion in $L_P^2$, obeys the relation
\begin{equation}
Tr [L_P^2 D_B^2] \propto L_P^{-2}\int d^4x\; \sqrt{g} \; R
\end{equation}
where $R$ is the Ricci scalar, and $L_P$ is Planck length. Thus the Dirac operator, by way of its eigenvalues [spectrum], captures the information of curvature algebraically. Thus, when the coordinates of the space-time are made into non-commuting operators, as we assume, this same Dirac operator stores the information of curvature, in our non-commutative matrix dynamics. In this non-commutative geometry, classical space-time is lost, but as a consequence of the Tomita-Takesaki theory, the new space possesses a one-parameter family of automorphisms, which play the role of a reversible time parameter, which we call Connes time, and denote as $\tau$.  

In the Planck scale matrix dynamics, we introduce the concept of an `atom' of space-time-matter [STM] ${q}$, this matrix being sum of a fermionic (matter) degree of freedom $q_F$, plus a bosonic degree of freedom $q_B$: ${ q} = { q_B} + { q_F}$. ${ q}$ is a matrix made of Grassmann elements, $q_B$ is Grassmann even, and $q_F$ is Grassmann odd. The fermionic part $q_F$ describes `matter' ($\dot{q}_F$),    and its associated Yang-Mills current ${ q_F}$. The bosonic part $q_B$ describes the fermion's associated Yang-Mills field $q_B$, and the associated space-time geometry $\dot{q}_B$. A dot denotes time-evolution with respect to Connes time $\tau$. Every STM atom has only one associated parameter, a length $L$, or equivalently, an area $L^2$. For example, an STM electron $q_e$ 
has a length $L_e$ (eventually to be identified with its Compton wave-length), and describes the electron and its current, and also the electron's gravitation and its Yang-Mills field. The universe at the Planck scale is made of  enormously many STM atoms, which all live in a Hilbert space ${\cal H}$ and evolve w.r.t.  time $\tau$. There are only two fundamental constants, the Planck length $L_P$ and Planck time $\tau_P$, these being units in which length and time are measured. There is no space-time at the Planck scale, nor a concept of mass or spin, or Planck's constant, all these being emergent at lower energies.

The action principle describing an STM atom is patterned after the action of a harmonic oscillator, and is given by
\begin{equation} 
\frac{S}{C_0} = \frac{1}{2}\int \frac{d\tau}{\tau_P} \; \frac{L_P^2}{L^4}\ Tr \left( \left [ \alpha q_B + L \dot{q}_B + \beta_1 \frac{L_P^2}{L^2}(\alpha q_F + L \dot{q}_F ) \right ] \times  \left [ \alpha q_B + L \dot{q}_B + \beta_2 \frac{L_P^2}{L^2}(\alpha q_F + L \dot{q}_F ) \right ] \right)
\label{acn}
\end{equation} 
$C_0$ is a constant with dimensions of action, $\alpha$ is the Yang-Mills coupling constant, and $\beta_1$ and $\beta_2$ are constant self-adjoint fermionic matrices. The operator Lagrangian inside the trace is bosonic. However, and this is crucial, neither the Lagrangian nor the action are real - they have a complex part, which is responsible for destroying superpositions in the classical limit. The choice of the action is motivated by the criterion that it should yield the standard Lagrangian of low-energy physics in the classical limit. Thus, the self-adjoint $\dot{q}_B$ is defined by the relation $cLD_B\equiv dq_B/d\tau$, where $D_B$ is the Dirac operator introduced above, and $c=L_P/\tau_P$. For the purpose of this essay, we restrict ourselves to fermions interacting with gravity, and ignore Yang-Mills fields (except to note that the latter imply entanglement), thus we set $\alpha=0$ and work with the action:
\begin{equation} 
\frac{S}{C_0} = \frac{1}{2}\int \frac{d\tau}{\tau_P}  \; \frac{L_P^2}{L^4}\ Tr \left( \left [L \dot{q}_B + \beta_1 \frac{L_P^2}{L^2}(L \dot{q}_F ) \right ] \times  \left [  L \dot{q}_B + \beta_2 \frac{L_P^2}{L^2}( L \dot{q}_F ) \right ] \right)
\end{equation} 
When there are many STM atoms, there is one such action term for each atom, and the total action is additive.

The two equations of motion can be obtained by varying w.r.t. $q_B$ and $q_F$ and show that the canonical momenta are constant:
\begin{align}
    p_B = \frac{\delta \textbf{L}}{\delta \dot{q}_B} &= \frac{a}{2}\bigg[2\dot{q}_B + \frac{L_P^2}{L^2}(\beta_1 +\beta_2)\dot{q}_F \bigg] = c_1\\ 
    p_F = \frac{\delta \textbf{L}}{\delta \dot{q}_F} &= \frac{a}{2} \frac{L_P^2}{L^2}\bigg[\dot{q}_B (\beta_1 +\beta_2)+ \frac{L_P^2}{L^2}\beta_1 \dot{q}_F \beta_2 +  \frac{L_P^2}{L^2}\beta_2 \dot{q}_F \beta_1 \bigg]=c_2
\end{align}
The first of these equations can be very usefully written as an eigenvalue equation
\begin{equation}
\left[D_B + \frac{L_P^2}{L^2}\frac{\beta_1+\beta_2}{2}D_F\right] \psi =  \frac{1}{L} \bigg(1+ i \frac{L_P^2}{L^2}\bigg)\psi
\label{eigen}
\end{equation}
where $Lc D_F \equiv dq_F/d\tau$. Although $D_B$ is self-adjoint, $D_F$ is not, and the imaginary part  of the eigenvalue is negligible so long as $L\gg L_P$. However, as we will see shortly, under certain circumstances the effective value of $L$ goes below $L_P$ and the imaginary part becomes significant. 

The above {\it deterministic}  matrix dynamics possesses a conserved charge of great significance, which is not there in ordinary classical dynamics, and which is the  result of a global unitary invariance of the trace Lagrangian.  This so-called Adler-Millard conserved charge is a matrix with the dimensions of action, and  is given by
\begin{equation}
    \tilde{C} = \sum_{r\in B}[q_r,p_r] -\sum_{r\in F} \{q_r,p_r \} 
    \label{amc}
\end{equation}
where the first sum is over all the bosonic matrices in the theory, and the second sum is over the fermionic ones. The sum over all commutators is conserved, even though each individual commutator is time-dependent. This is where Planck's constant will emerge from, in the low-energy limit.

We now ask, what does the average motion of one or more STM atoms look like, if we are not observing them over Planck time resolution, but over much bigger time intervals? This is equivalent to asking what the mean matrix dynamics is, at low energies. It is possible to find the answer using the conventional techniques of statistical mechanics, and find the thermodynamic equilibrium state. One considers an ensemble of many copies of the matrix dynamics system, each of them operating at the Planck scale. A probability density of their distribution in phase space is defined, and the Boltzmann entropy constructed from it is extremised, subject to the conservation of the Adler-Millard charge, and conservation of energy. We also take the low energy limit, equivalent to assuming that coarse-graining has been done over time-intervals much larger than Planck time.

The resultant mean dynamics falls in one of two limiting cases, or a mix thereof. The first limit is when each of the STM atoms has a length scale $L$ such that $L\gg L_P$. Then the imaginary part of the eigenvalue in Eqn. (\ref{eigen}) can be neglected: this is equivalent to the Hamiltonian of the theory being self-adjoint, with the anti-self-adjoint part being ignorable. Under this important approximation and assumption, the emergent  mean dynamics in this limit is quantum general relativity. The conserved Adler-Millard charge is equipartitioned over all degrees of freedom, with each commutator and anti-commutator now the same constant, which is identified with Planck's constant $\hbar$. 
Evolution is still w.r.t. Connes time $\tau$ (no apace-time yet), and this emergent theory comes into play  if no background space-time is available even at low energies (e.g. what is the gravitational field of an electron in the double-slit interference experiment, after crossing the slits, but before reaching the screen). Each of the bosonic and fermiionic degrees of freedom obeys Heisenberg equations of motion, which is equivalent to a functional Schrodinger evolution driven by Connes time (a resolution of the problem of time in quantum general relativity).

Now consider the other extreme limit, in which a very large number of STM atoms undergo  entanglement, caused by their 
Yang-Mills interactions [strong, electroweak]. The effective length $L_{eff}$ associated with the entangled system is reduced by a factor $1/N$, where $N$ is the number of entangled STM atoms. For sufficiently large $N$, $L_{eff}$ goes below Planck length, the anti-self-adjoint part of the Hamiltonian becomes important, and the equilibrium approximation leading to the emergence of deterministic quantum theory breaks down. From the viewpoint of the mean dynamics at energies below Planck scale, the imaginary part of the Hamiltonian represents rapid imaginary variations. Because the underlying deterministic Planck scale dynamics has been coarse-grained, and we are only viewing an averaged dynamics, these rapid variations act as random fluctuations which disturb the mean quantum dynamics significantly. This is the equivalent of the random kicks of the pollen grain in a glass of water. These rapid imaginary variations are modelled as a stochastic anti-Hermitean noise, which then destroys quantum superpositions (over a certain time-scale). This results in the so-called spontaneous localisation of (only the fermionic part) of the STM atoms. It is akin to transiting from a matrix-valued description of dynamics (many eigenvalues included at the same time) to a c-number valued description (classical dynamics, only one eigenvalue). The eigenvalues of $q_F$ to which the fermionic degrees of freedom localise play the role of space-time coordinates, and that is how the space-time manifold emerges. The bosonic part of the STM atom plays the role of the gravitational field `produced' by the spontaneously localised fermionic part (material bodies). Because the fermionic source does not obey position superpositions in this limit, neither are the corresponding space-time geometries superposed.
We have shown earlier \cite{Maithresh2019} that spontaneous localisation of such a large number of entangled STM atoms transforms their total action into the action for general relativity, with Newton's gravitational constant $G$ defined in an emergent sense as 
$G\equiv L_P^2 c^3 / \hbar$:
\begin{equation}
S_{total} = \int d\tau \sum_i Tr D^2_i  \quad  {\mathbf \longrightarrow} \quad \int d\tau \bigg[ \frac{c^3}{2G}\int d^4x \; \sqrt{g} \; R + \int d^4x \; \sqrt{g} \; c\; \sum_i m_i \delta^3({\bf x} - {\bf x_0})\bigg]
\end{equation}
where $D_i^2$ is a condensed notation for the operator Lagrangian of a single STM atom, as in (\ref{acn}) above.

Those STM atoms which have not undergone localisation, can be described via the original matrix dynamics. Or, as is conventionally done, their gravitational aspect can be neglected, and the fermionic part can be described as a quantum field theory on the emergent space-time background.

We can say that: Deterministic Matrix Dynamics on Planck scale $\longrightarrow$ at lower energies $\longrightarrow$ Quantum Gravity + Spontaneous Localisation. In conventional approaches to quantum gravity, the spontaneous localisation part is missing. Because we allowed for a matrix dynamics more general than quantum theory, and allowed the fundamental Hamiltonian to have a non-self-adjoint part, and coupled gravity to Dirac fermions, we get the classical limit in a satisfactory way.

We can now also explain, by addressing the notorious quantum measurement problem, why the wave function of the quantum system collapses, randomly, to one of the eigenstates of the observable being measured. The time of arrival of the quantum system at the measuring apparatus is crucial, down to Planck time resolution. For, once the interaction of the quantum system with the apparatus begins, further evolution depends on the rapid imaginary variations taking place in the Planck scale matrix dynamics of the combined system. These variations are important now, since an enormous number of STM atoms have gotten entangled with the quantum system. Which way the full system will go can be predicted if and only if one is examining the dynamics on Planck time scales. That however, is possible only when Planck scale energies are used in the probe; and impossible under current laboratory conditions. As a result, because the measurement is not precise enough, one does not know at which stage of the rapid variation did the quantum system actually hit the apparatus. The outcome of the measurement hence appears random. But, truly speaking, Nature does not play Dice at the Planck scale. It only appears to play dice, at lower energies, because our probes are not precise enough. Thus, the measurement problem is solved by an underlying quantum theory of gravity at the Planck scale.The Schrodinger equation is deterministic, and the evolution of the quantum system during measurement is also deterministic. Deterministic, but non-local! Bell's inequalities hold.

An important by-product of our theory is an explanation of the inferred dark-energy as an infra-red quantum gravitational phenomenon \cite{Singh:DE}. The imaginary variations in length, as present in the above Eqn. (\ref{eigen}), imply a minimum uncertainty 
$\Delta L$  in length measurements, using a probe \cite{SinghqgV2019}:
\begin{equation}
(\Delta L)^3 \sim L_P^2 \; L
\end{equation}
This so-called Karolyhazy uncertainty relation has been known in the literature, but our work provides the first {\it ab\ initio} derivation for it, from quantum gravity. Note that this uncertainty is larger than Planck length by a factor $L/L_P$: the uncertainty is not absolute and universal, but depends on the length being measured. This has powerful implications, including that the information content in a spatial region increases holographically. Because, considering a region of volume $L^3$, it can be divided into fundamental cells of volume $(\Delta L)^3$, with $\Delta L$ as given above. Thus the number of information units in this region is $L^3 / (\Delta L)^3\sim (L/L_P)^2$, which of course is a holographic property: information content increases not as volume, but as area.

This has implications for cosmology. Taking $L\sim 10^{28}$ cm for the observed universe implies there are approximately $10^{122}$ units of information in the Hubble volume. But this is puzzling: there are only some $10^{79}$ particles in the observed universe, pushed up to about $10^{90}$ or so, if dark matter is included. Where then are these $10^{122}$ degrees of freedom? We propose that these degrees of freedom come from a class of extremely light `dark energy particles' which have no interaction other than gravitational, have a mass of the order of $10^{-33}$ eV/c$^2$ each,  and hence a Compton length of the order of Hubble radius. Because of this enormous wavelength, they are frozen in, and have very negligible kinetic energy. Their only associated energy is gravitational, and moreover, because they have no Yang-Mills interaction, they remain unentangled and behave just like dark energy. Being ultra-light means their time scale for spontaneous collapse is enormous - they are not at all classical, but are uncollapsed quantum gravitational STM atoms. These quantum gravitational entities dominate the universe, and are causing the classical spontaneously collapsed material universe to accelerate. Their existence is evidence for quantum gravity, and the physics that justifies their presence is the same physics that explains why a chair is never seen in two places at the same time!

\bigskip

\centerline{\bf REFERENCES}

\bibliography{biblioqmtstorsion}

\end{document}